\documentclass[runningheads]{llncs}
\usepackage[T1]{fontenc}
\usepackage{graphicx}
\usepackage{hyperref}
\usepackage{color}

\usepackage{listings}
\usepackage{amsfonts} 
\usepackage{booktabs, colortbl}
\usepackage[dvipsnames]{xcolor}
\lstdefinestyle{mystyle} {
	aboveskip=20pt,
	keywordstyle=\color{blue},
	commentstyle=\itshape\color{purple},
	stringstyle=\color{PineGreen},
	basicstyle=\ttfamily\footnotesize,
	breakatwhitespace=false,         
	breaklines=true,                 
	captionpos=b,       
	keepspaces=true,
	showspaces=false,                
	showstringspaces=false,
	showtabs=false,                  
	tabsize=2
}
\definecolor{lavender}{rgb}{0.9,0.9,0.98}

\usepackage{array}
\newcolumntype{L}{>{\centering\arraybackslash}m{0.14\textwidth}}

\newcolumntype{K}{>{\centering\arraybackslash}m{0.2\textwidth}}

\newcolumntype{P}[1]{>{\centering\arraybackslash}p{#1}}

\usepackage{comment,cite}
\usepackage{subcaption} 
\usepackage{wrapfig}

\usepackage{todonotes}
\newcommand{\knote}[1]{\todo[inline, color=blue!20]{#1}}

\begin{document}
\title{Neural Networks in Imandra: Matrix Representation as a Verification Choice\thanks{E.Komendantskaya acknowledges support of EPSRC grant EP/T026952/1.}}
%
%
\author{Remi Desmartin\inst{1} \and
Grant Passmore\inst{2}\and
Ekaterina Kommendentskaya\inst{1}}
\authorrunning{R. Desmartin et al.}
%
\institute{
	Heriot-Watt University, Edinburgh, UK
	\email{\{rhd2000,e.komendantskaya\}@hw.ac.uk}	
\and
	Imandra, Austin TX, USA
	\email{grant@imandra.ai}
	\url{http://www.imandra.ai}
}
\maketitle              
\begin{abstract}
The demand for formal verification tools for neural networks has increased as
neural networks have been deployed in a growing number of safety-critical
applications. Matrices are a data structure essential to formalising neural
networks. Functional programming languages encourage diverse approaches to
matrix definitions. This feature has already been successfully exploited in
different applications. The question we ask is whether, and how, these ideas can
be applied in neural network verification. A functional programming language
Imandra combines the syntax of a functional programming language and the power
of an automated theorem prover. Using these two key features of Imandra, we
explore how different implementations of matrices can influence automation of
neural network verification.

\end{abstract}

\section{Motivation}

Neural network (NN) verification was pioneered by the
SMT-solving~\cite{KaBaDiJuKo17Reluplex,HuangKWW17} and an abstract
interpretation~\cite{SinghGPV19,GeMiDrTsChVe18,AEHW20} communities. However, recently claims have been made that functional programming, too, can be valuable in this domain. 
 There is a library~\cite{MariaBLFGRG22}
formalising small rational-valued neural networks in Coq. A more sizeable
formalisation called MLCert~\cite{BS19} imports neural networks from Python, treats
floating point numbers as bit vectors, and proves properties describing the
generalisation bounds for the neural networks.
An $F^*$ 
formalisation~\cite{KokkeKKAA20} uses $F^*$ reals and
refinement types for proving the robustness of networks trained in Python.

There are several options for defining neural networks in functional
programming, ranging from defining neurons as record types~\cite{MariaBLFGRG22}
to treating them as functions with refinement types~\cite{KokkeKKAA20}. But we
claim that two general considerations should be key to any NN formalisation
choice. Firstly, we must define neural networks as executable
functions, because we want to take advantage of executing them in the functional
language of choice. Secondly, a generic approach to layer definitions is needed,
particularly when we implement complex neural network architectures, such as
convolutional layers.

These two essential requirements dictate that neural networks are represented as
matrices, and that a programmer makes choices about matrix formalisation. This
article will explain these choices, and the consequences they imply,
from the verification point of view. We use Imandra~\cite{PassmoreCIABKKM20} to
make these points, because Imandra is a functional programming language with
tight integration of automated proving.

Imandra has been successful as a user-friendly and scalable tool in the FinTech
domain~\cite{Passmore21}. The secret of its success lies in a combination of
many of the best features of functional languages and interactive and automated
theorem provers. Imandra's logic is based on a pure, higher-order subset of
OCaml, and functions written in Imandra are at the same time valid OCaml code
that can be executed, or \emph{``simulated''}. Imandra's mode of interactive
proof development is based on a typed, higher-order lifting of the
\emph{Boyer-Moore waterfall}~\cite{BM79} for automated induction, tightly
integrated with novel techniques for SMT modulo recursive functions.

This paper builds upon the recent development of a CheckINN, a NN verification library in Imandra~\cite{DPKD22}, but
discusses specifically the matrix representation choices and their consequences.

\section{Matrices in Neural Network Formalisation}
We will illustrate the functional approach to neural network formalisation and
will introduce the syntax of the Imandra programming
language~\cite{PassmoreCIABKKM20} by means of an example. When we say we want to
formalise neural networks as functions, essentially, we aim to be able to define
a NN using just a line of code:

\begin{lstlisting}[language=caml, label={lst:model}]
  let cnn input =
      layer_0 input >>= layer_1 >>= layer_2 >>= layer_3
\end{lstlisting}

where each \lstinline?layer_i? is defined in a modular fashion.

 To see that a functional approach to neural networks does not necessarily imply generic nature of the code, 
let us consider an example. 
A \emph{perceptron}, also known as a \emph{linear classifier}, classifies a given input vector $X = (x_1, ..., x_m)$ into one of two classes $c_1$ or $c_2$ by computing a linear combination of the input vector with a vector of synaptic weights $(w_0, w_1, ..., w_m)$, in which $w_0$ is often called an \emph{intercept} or \emph{bias}: 
	$ f(X) = 	\sum_{i=1}^{m}w_ix_i + w_0 $.
If the result is positive, it classifies the input as $c_1$ and if negative as $c_2$. It effectively divides the input space along a hyperplane defined by
$\sum_{i=1}^{m}w_ix_i + w_0 = 0$. 



In most classification problems, classes are not linearly separated. To handle such problems, we can apply a non-linear function $a$ called an \textit{activation function} to the linear combination of weights and inputs. The resulting definition of a perceptron $f$ is:
\begin{equation}\label{eq:perceptron}
	f(X) = a\left(\sum_{i=1}^{m}w_ix_i + w_0\right)
      \end{equation}

     Let us start with a naive prototype of perceptron in Imandra.  The Iris data set is a ``Hello World'' example in data mining; it represents 3 kinds of Iris flowers using 4 selected features.
      In Imandra, inputs can be represented as a data type: 

\begin{lstlisting}[language=caml]
type iris_input = {
  sepal_len: real;
  sepal_width: real;
  petal_len: real;
  petal_width: real;}
\end{lstlisting}

And we define a perceptron as a function:

\begin{lstlisting}[language=caml]
let layer_0 (w0, w1, w2, w3, w4) (x1, x2, x3, x4) =
  relu (w0 +. w1 *. x1 +. w2 *. x2 +. w3 *. x3 +. w4 *. x4)
\end{lstlisting}
where \lstinline{*.} and   \lstinline{+.} are \emph{times} and \emph{plus} defined on reals.  Note the use of the \lstinline{relu} activation function, which returns $0$ for all negative inputs and acts as the identity function otherwise.

Already in this simple example, one perceptron is not sufficient, as we must map its output to three classes. We use the usual machine learning literature trick and define a further layer of $3$ neurons, each representing one class. Each of these neurons is itself a perceptron, with one incoming weight and one bias. This gives us:

\begin{lstlisting}[language=caml]
let layer_1 (w1, b1, w2, b2, w3, b3) f1 =
  let o1 = w1 *. f1 +. b1 in
  let o2 = w2 *. f1 +. b2 in
  let o3 = w3 *. f1 +. b3 in
  (o1, o2, o3)

let process_iris_output (c0, c1, c2) =
  if (c0 >=. c1) && (c0 >=. c2) then "setosa"
  else if (c1 >=. c0) && (c1 >=. c2) then "versicolor"
  else "virginica"
\end{lstlisting}

\noindent The second function maps the output of the three neurons to the three specified classes. This post-processing stage often takes the
form of an \emph{argmax} or \emph{softmax} function, which we omit. 

And thus the resulting function that defines our neural network model is:


\begin{lstlisting}[language=caml]
let model input = process_iris_input input
        |> layer_0 weights_0 |> layer_1 weights_1 |>
                                        process_iris_output
\end{lstlisting}

Although our naive formalisation has some features that we desired from the start, i.e.\ it defines a neural network as a composition of functions,
it is too inflexible to work with arbitrary compositions of layers.
In neural networks with hundreds of weights in every layer this manual approach will quickly become infeasible (as well as error-prone).
So, let us generalise this attempt from the level of individual neurons to the level of matrix operations.

The composition of many perceptrons is often called a \emph{multi-layer perceptron (MLP)}.
An MLP consists of an input vector (also called input layer in the literature),
multiple hidden layers and an output layer, each layer 
made of perceptrons with weighted connections to the previous layers' outputs. 
The weight and biases of all the neurons in a layer can be represented by two matrices denoted by $W$ and $B$. By adapting equation \ref{eq:perceptron} to this matrix notation, a layer's output $L$ can be defined as:
\begin{equation}
	L(X) = a(X \cdot W + B)
\end{equation}
where the operator $\cdot $ denotes the dot product between $X$ and each row of $W$, 
$X$ is the layer's input and $a$ is the activation function shared by all nodes in a layer. 
As the dot product multiplies pointwise all inputs by all weights, such layers are often called \emph{fully connected}. 

By denoting $a_k, W_k, B_k$ --- the activation function, weights and biases of the $k$th layer respectively, an MLP $F$ with $L$ layers is traditionally defined as:
\begin{equation}
	F(X) = a_L[B_L + W_L (a_{L-1}(B_{L-1} + W_{L-1}(...(a_1(B_1+W_1\cdot X)))))]
      \end{equation}

      At this stage, we are firmly committed to using matrices and matrix operations. And we have two key choices:
\begin{enumerate}
\item to represent matrices as lists of lists (and take advantage of the inductive data type \lstinline{List}),
\item define matrices as functions from indices to matrix elements,
\item or take advantage of record types, and define matrices as records with maps.  
\end{enumerate}

The first choice was taken in \cite{heras_incidence_2011} (in the context of dependent types in Coq), in ~\cite{KokkeKKAA20} (in the context of refinement types of F$^*$) and in \cite{grant_sparse_1996} (for sparse matrix encodings in Haskell).  The difference between the first and second approaches was
discussed in~\cite{wood_vectors_2019} (in Agda, but with no neural network application in mind).
The third method was taken in ~\cite{MariaBLFGRG22} using Coq  (records were used there to encode individual neurons).

In the next three sections, we will systematise these three approaches using the same formalism and the same language,
and trace the influence of these choices on neural network verification.

\section{Matrices as Lists of Lists} \label{sec:lists}
We start with re-using  Imandra's  \lstinline{List} library. Lists are defined as inductive data structures:

\begin{lstlisting}[language=caml]
type 'a list =
| []
| (::) of 'a * 'a list
\end{lstlisting}

Imandra holds a comprehensive library of list operations covering a large part of OCaml's standard \lstinline{List} libary, which we re-use in the definitions below.
We start with defining vectors as lists, and matrices as lists of vectors. 

\begin{lstlisting}[frame=none, language=caml]
type 'a vector = 'a list
type 'a matrix = 'a vector list 
\end{lstlisting}

It is possible to extend this formalisation by using dependent~\cite{heras_incidence_2011} or refinement~\cite{KokkeKKAA20} types to check the matrix size, e.g. when performing matrix multiplication. But in Imandra this facility is not directly available, and we will need to use error-tracking (implemented via the monadic \lstinline{Result} type) to facilitate checking of the matrix sizes.

As there is no built-in type available for matrices equivalent to \lstinline{List} for vectors, the \lstinline{Matrix} module implements a number of functions for basic operations needed throughout the implementation. For instance, \lstinline{map2} takes as inputs a function $f$ and two matrices $A$ and $B$ of the same dimensions and outputs a new matrix $C$ where each element $c_{i,j}$ is the result of $f(a_{i, j}, b_{i, j})$:

\begin{lstlisting}[frame=none, language=caml]
  let rec map2 (f:'a -> 'b -> 'c) (x:'a matrix) (y:'b matrix) =
    match x with
	  | [] -> begin match y with
   		      | [] -> Ok []
		        | y::ys  -> Error "map2: invalid length." end
 	  | x::xs -> begin match y with 
  		         | [] -> Error "map2: invalid length." 
		           | y::ys -> let hd = map2 f x y in
		                      let tl = map2 f xs ys in
		                      lift2 cons hd tl end
\end{lstlisting}

This implementation allows us to define other useful functions concisely. For instance, the dot-product of two matrices 
is defined as:

\begin{lstlisting}[frame=none, language=caml]
let dot_product (a:real matrix) (b:real matrix): ('a, real matrix) result =
	Result.map sum (map2 ( *. ) a b)
\end{lstlisting}

Note that since the output of the function \lstinline{map2} is wrapped in the monadic \lstinline{result} type, we must use \lstinline{Result.map} to apply \lstinline|sum|. Similarly, we use standard monadic operations for the \lstinline|result| monad such as \lstinline|bind| or \lstinline|lift|.


A fully connected layer is then defined as a function \lstinline{fc} that takes
as parameters an activation function, a 2-dimensional matrix of
layer's weights and an input vector:
\begin{lstlisting}[caption=Fully connected layer implementation, language=caml, label={lst:fully_connected}]
let activation f w i = (* activation func., weights, input *)
 let linear_combination m1 m2 = if (length m1) <> (length m2)
     then Error "invalid dimensions" 
     else map sum (Vec.map2 ( *. ) m1 m2) in
 let i' = 1.::i in (* prepend 1. for bias *)
 let z = linear_combination w i' in
 map f z
	
let rec fc f (weights:real matrix) (input:real vector) =
 match weights with
 | [] -> Ok []
 | w::ws -> lift2 cons (activation f w input) (fc f ws input)
\end{lstlisting}

Note that each row of the weights matrix
represents the weights for one of the layer's nodes. The bias for each node is
the first value of the weights vector, and $1$ is prepended to the input vector
when computing the dot-product of weights and input to account for that.

  It is now easy to see that our desired modular approach to composing layers works as stated. We may define the layers using the syntax:
  \lstinline{let layer_i = fc a weights}, where \lstinline{i} stands for \lstinline{0,1,2,3}, and \lstinline{a} stands for any chosen activation function. 

  Although natural, this formalisation of layers and networks suffers from two problems.  Firstly, it lacks the matrix dimension checks that were readily provided  via refinement types in~\cite{KokkeKKAA20}. This is because Imandra is based on a computational fragment of HOL, and has no refinement or dependent types. To mitigate this, our library performs explicit dimension checking via a {\tt result} monad, which clutters the code and adds additional computational checks.
  Secondly, the matrix definition via the list datatypes makes verification of neural networks very inefficient.
  This general effect has been already reported in \cite{KokkeKKAA20}, but it may be instructive to look into the problem from the Imandra perspective.

   Robustness of neural networks~\cite{CKDKKAE22} is best amenable to proofs by arithmetic manipulation. This explains the interest of the SMT-solving community in the topic, which started with using Z3 directly~\cite{HuangKWW17}, and has resulted in highly efficient SMT solvers specialised on robustness proofs for neural networks~\cite{KaBaDiJuKo17Reluplex,KatzHIJLLSTWZDK19}.   Imandra's waterfall method~\cite{PassmoreCIABKKM20} defines a default flow for the proof search, which starts with unrolling inductive definitions, simplification and rewriting.
  As a result, proofs of neural network robustness or proofs as in the ACAS Xu challenge~\cite{KaBaDiJuKo17Reluplex,KatzHIJLLSTWZDK19}, which do not actually need induction,
  are not efficiently tackled using Imandra's inductive waterfall: the proofs simply do not terminate.

There is another verification approach available in Imandra which is better suited for this type of problem:  \lstinline{blast}, a tactic for SAT-based symbolic execution modulo
higher-order recursive functions. Blast is an internal custom SAT/SMT solver that can be called explicitly to discharge an Imandra verification goal.
 However, \lstinline{blast} currently does not support real arithmetic. This
 requires us to \emph{quantize} the neural networks we use (i.e.\ convert them to
 integer weights) and results in a \emph{quantised NN implementation}~\cite{DPKD22}.
 However, even with quantisation and the use of  \lstinline{blast}, while we succeed on many smaller benchmarks, Imandra fails to scale `out of the box' to the ACAS Xu challenge, let alone larger neural networks used in computer vision.

 This also does not come as a surprise: as~\cite{KaBaDiJuKo17Reluplex} points out, general-purpose SMT solvers do not scale to NN verification challenges.
This is why the algorithm \lstinline{reluplex} was introduced in \cite{KaBaDiJuKo17Reluplex} as an additional heuristic to SMT solver algorithms;
\lstinline{reluplex} has since given rise to a domain specific solver Marabou~\cite{KatzHIJLLSTWZDK19}.
Connecting Imandra to Marabou may be a promising future direction. 

However, this method of matrix formalisation can still bring benefits. When
we formulate verification properties that genuinely require induction,
formalisation of matrices as lists does result in more natural, and easily
automatable proofs. For example, De Maria et al.~\cite{MariaBLFGRG22} formalise
in Coq \emph{``neuronal archetypes''} for biological neurons. Each archetype is
a specialised kind of perceptron, in which additional functions are added to
amplify or inhibit the perceptron's outputs. It is out of the scope of this paper to
formalise the neuronal archetypes in Imandra, but we take methodological insight
from~\cite{MariaBLFGRG22}. In particular, De Maria et al. show that there
are natural higher-order properties that one may want to verify.

To make a direct comparison, modern neural network
verifiers~\cite{KaBaDiJuKo17Reluplex,SinghGPV19} deal with verification tasks of
the form ``given a trained neural network $f$, and a property $P_1$ on its
inputs, verify that a property $P_2$ holds for $f$'s outputs''. However, the
formalisation in~\cite{MariaBLFGRG22} considers properties of the form ``any
neural network $f$ that satisfies a property $Q_1$, also satisfies a property
$Q_2$.'' Unsurprisingly, the former kind of properties can be resolved by
simplification and arithmetic, whereas the latter kind requires
induction on the structure of $f$ (as well as possibly nested induction on
parameters of $Q_1$).

Another distinguishing consequence of this approach is that it is orthogonal to
the community competition for scaling proofs to large networks: usually the
property $Q_1$ does not restrict the size of neural networks, but rather points
to their structural properties. Thus, implicitly we quantify over neural
networks of any size.

To emulate a property \emph{\`a la} de Maria et al.,  in~\cite{DPKD22} we  defined a general network monotonicity property: \emph{any fully connected
network with positive weights is \emph{monotone}, in the sense that, given
increasing positive inputs, its outputs will also increase}. There has been some
interest in monotone networks in the literature~\cite{JS98,WehenkelL19}.
Our experiments show that Imandra can prove such properties by induction on the networks'
structure almost automatically (with the help of a handful of auxiliary lemmas). And the proofs easily go through for both quantised and real-valued neural networks.\footnote{Note that in these experiments, the implementation of weight matrices as lists of lists is implicit -- we redefine matrix manipulation functions that are less general but more convenient for proofs by induction.}

\section{Matrices as Functions} \label{sec:function}

We now return to the verification challenge of ACAS Xu, which we failed to
conquer with the inductive matrix representation of the last section. This time
we ask whether representing matrices as functions and leveraging Imandra's
default proof automation can help.

With this in mind, we redefine matrices as functions from indices to values, which gives
constant-time (recursion-free) access to matrix elements:

\begin{lstlisting}[
	caption=Implementation of matrices as functions from indices to values,
	label={lst:function},
	language=caml
	]
type arg =
  | Rows
  | Cols
  | Value of int * int

type 'a t = arg -> 'a
\end{lstlisting}

Note the use of the \lstinline{arg} type, which treats a matrix as a function evaluating ``queries'' (e.g., ``how many rows does this matrix have?'' or ``what is the value at index $(i,j)$?''). This formalisation technique is used as Imandra's logic does not allow function values inside of algebraic data types. 
We thus recover some functionality given by refinement types in~\cite{KokkeKKAA20}.  

Furthermore, we can map functions over a matrix or a pair of matrices  (using \lstinline{map2}), transpose a
matrix, construct a diagonal matrix etc.\ without any recursion, since we work
point-wise on the elements. At the same time, we remove the need for error
tracking to ensure matrices are of the correct size: because our matrices are
total functions, they are defined everywhere (even outside of their stated
dimensions), and we can make the convention that all matrices we build are valid
and sparse by construction (with default 0 outside of their dimension bounds).

The resulting function definitions are much more succinct than with lists of lists; take for instance \lstinline{map2}:

\begin{lstlisting}[language=caml]
let map2 (f: 'a -> 'b -> 'c) (m: 'a t) (m': 'b t) : 'c t =
	function
		| Rows -> rows m
		| Cols -> cols m
		| Value (i,j) -> f (m (Value (i,j))) (m' (Value (i,j)))
\end{lstlisting}

\noindent This allows us to define fully connected layers:

\begin{lstlisting}[language=caml]
let fc (f: 'a -> 'b) (weights: 'a Matrix.t) (input: 'a Matrix.t) = 
	let open Matrix in
	function
		| Rows         -> 1
		| Cols         -> rows weights
		| Value (0, j) -> 
			let input' = add_weight_coeff input in
			let weights_row = nth_row weights j in
			f (dot_product weights_row input')
		| Value _      -> 0
\end{lstlisting}

\noindent As the biases are included in the \lstinline|weights| matrix, \lstinline|add_weight_coeff| prepends a column with coefficients $1$ to the input so that they are taken into account.

For full definitions of matrix operations and layers, the reader is referred to~\cite{DPKD22}, but we will give some definitions here, mainly to convey the general style (and simplicity!) of the code.  Working with the ACAS Xu networks~\cite{KaBaDiJuKo17Reluplex}, a script transforms the original networks into sparse functional matrix representation.
For example, layer 5 of one of the networks we used is defined as follows:

\begin{lstlisting}[language=caml]
let layer5 = fc relu (
  function
  | Rows -> 50
  | Cols -> 51
  | Value (i,j) -> Map.get (i,j) layer5_map)

let layer5_map =
  Map.add (0,0) (1) @@
  Map.add (0,10) (-1) @@
  Map.add (0,29) (-1) @@
  ...
  Map.const 0
\end{lstlisting}

\noindent The sparsity effect is achieved by \emph{pruning} the network, i.e. removing weights that have the smallest impact on the network's performance. The weight's magnitude is used to select those to be pruned. This method, though rudimentary, is considered a reasonable pruning technique \cite{lecun_optimal_1989}.
%
We do this mainly in order to reduce the amount of computation Imandra needs to perform, and to make the verification problem amenable to Imandra.

With this representation, we are able to verify the properties described in ~\cite{KaBaDiJuKo17Reluplex} on some of the pruned networks (see Table~\ref{tab:acas_xu}).
This is a considerable improvement compared to the previous section, where the implementation did not allow to verify even pruned networks. It is especially impressive that it comes ``for free'' by simply changing the underlying matrix representations.

Several factors played a role in automating the proof.
Firstly, by using maps for the
large matrices, we eliminate all recursion (and large case-splits) except for
matrix folds (which now come in only via the dot product), which allowed Imandra
to expand the recursive matrix computations on demand.
Secondly, Imandra's native simplifier contributed to the success. It works on a
DAG representation of terms and speculatively expands instances of recursive
functions, only as they are (heuristically seen to be) needed. Incremental
congruence closure and simplex data structures are shared across DAG nodes, and
symbolic execution results are memoised. The underlying \lstinline{Map.t}
components of the functions are reasoned about using a decision procedure for
the theory of arrays.
Informally speaking, Imandra works lazily expanding out the linear algebra as it is needed, and eagerly sharing information over the DAG.
Contrast this approach with that of Reluplex which, informally, starts with the linear algebra fully expanded, and then works to derive laziness and sharing. 


Although Imandra's simplifier-based automation above could give us results which
\lstinline{blast} could not deliver for the same network, it still did not scale
to the original non-quantised (dense) ACAS Xu network. Contrast this with
domain-specific verifiers such as Marabou which are able to scale (modulo
potential floating point imprecision) to the full ACAS Xu.
We are encouraged that the results of
this section were achieved without tuning Imandra's generic proof automation
strategies, and hopeful that the development of neural-network specific tactics 
will help Imandra
scale to such networks in the future. 

\begin{table}\label{tab:acas_xu}
	\centering
	\begin{tabular}{{ P{1.5cm}  P{2cm}  P{2cm} P{1.5cm}  P{2cm} P{1.5cm} }}
		\toprule
		\multicolumn{2}{P{3.6cm}}{} & \multicolumn{2}{P{3.6cm}}{\textbf{CheckINN: Pruned Networks}} & \multicolumn{2}{P{3.6cm}}{\textbf{Reluplex: Full ACAS Xu Networks}} \\
		\cmidrule(rl){3-4} \cmidrule(rl){5-6}
		\textbf{Property} & \textbf{Result} & Quantity & Time (s) & Quantity & Time (s) \\ 
		\midrule
		\rowcolor{lavender}
		$\phi1$ & SAT 		& 20 & 13387 	& 0  &			\\
		\rowcolor{lavender}
		& UNSAT 	& 0	 & 		 	& 41 & 394517	\\
		\rowcolor{lavender}
		& TIMEOUT 	& 24 & 			& 4  &			\\
		$\phi2$ & SAT 		& 7  & 2098 	& 35 & 82419	\\
		& UNSAT 	& 2	 & 896	 	& 1  & 463		\\
		& TIMEOUT 	& 26 & 			& 4  &			\\
		\rowcolor{lavender}
		$\phi3$ & SAT 		& 39 & 10453 	& 35 & 82419	\\
		\rowcolor{lavender}
		& UNSAT 	& 0	 & 		 	& 1  & 463		\\
		\rowcolor{lavender}
		& TIMEOUT 	& 2  & 			& 4  &			\\
		$\phi4$ & SAT 		& 36 & 21533	& 0  & 		 	\\
		& UNSAT 	& 0  & 			& 32 & 12475 	\\
		& TIMEOUT 	& 5  & 			& 0  &			\\
		\rowcolor{lavender}
		$\phi5$ & SAT 		& 1  & 98	 	& 0  & 			\\
		\rowcolor{lavender}
		& UNSAT 	& 0  & 		 	& 1	 & 19355	\\
		$\phi6$ & SAT 		& 1  & 98		& 0  &			\\
		& UNSAT 	& 0  & 			& 1  & 180288	\\
		\rowcolor{lavender}
		$\phi7$ & TIMEOUT 	& 1  &	 		& 1  & 			\\
		$\phi8$ & SAT 		& 0  & 			& 1  & 40102 	\\
		& TIMEOUT 	& 1  & 			& 0  & 			\\
		\rowcolor{lavender}
		$\phi9$ & SAT		& 1  & 109	 	& 0  & 			\\
		\rowcolor{lavender}
		& UNSAT		& 0  & 			& 1  & 99634 	\\
		$\phi10$ & SAT		& 0  &			& 0  &			\\
		& UNSAT 	& 1  & 637 		& 1  & 19944	\\
		& TIMEOUT 	& 0  &	 		& 0 &			\\
		\bottomrule
	\end{tabular}
	\caption{Results of experiments ran on the properties and networks from the ACAS Xu benchmark~\cite{KaBaDiJuKo17Reluplex}. The CheckINN verifications were run with 90\% of the weights pruned, on virtual machines with four 2.6 GHz Intel Ice Lake virtual processors and 16GB RAM. Timeout was set to 5 hours}
\end{table}

\section{Real-valued Matrices; Records  and Arrays}\label{sec:records}

It is time we turn to the question of incorporating real values into matrices.
Section~\ref{sec:lists} defined matrices as lists of lists; and that definition
in principle worked for both integer and real-valued matrices. However, we could
not use \lstinline{[@@blast]} to automate proofs when real values were involved;
this meant we were restricted to verifying integer-valued networks. On the
other hand, the matrix as function implementation extends to proofs with real
valued matrices, however it is not a trivial extension. In the functional
implementation, the matrix's value must be of
the same type as its dimensions (Listing~\ref{lst:function}). Thus, if the matrix elements are real-valued, then
in this representation the matrix dimensions will be real-valued as well. This,
it turns out, is not trivial to deal with for functions which do recursion along
matrix dimensions.


To simplify the code and the proofs, three potential solutions were considered:
\begin{itemize}
	\item Using an algebraic data type for results of matrix queries: this introduces pattern matching in the implementation of matrix operations, which Section~\ref{sec:lists} taught us to avoid.
	\item Define a matrix type with real-valued dimensions and values: this poses the problem of proving the function termination when using matrix dimensions in recursion termination conditions.
	\item Use \emph{records} to provide polymorphism and allow matrices to use integer dimensions and real values.
\end{itemize}

This section focuses on these three alternatives.

\subsection{Algebraic Data Types for  Real-Valued Matrices}

The first alternative is to introduce an algebraic data type that allows the matrix functions to return either reals or integers. 

\begin{lstlisting}[language=caml]
type arg =
	| Rows
	| Cols
	| Value of int * int
	| Default

type 'a res = 
	| Int of int
	| Val of 'a

type 'a t = arg -> 'a res
\end{lstlisting}

This allows a form of polymorphism, but it also introduces pattern matching each time we query a value from the matrix. For instance, in order to use dimensions as indices to access a matrix element we have to implement the following \lstinline{nth_res} function:  

\begin{lstlisting} [language=caml]
let nth_res (m: 'a t) (i: 'b res) (j: 'c res): 'a res = match (i, j) with 
	| (Int i', Int j') -> m (Value (i', j'))
	| _                -> m Default
\end{lstlisting}

The simplicity and efficiency of the functional implementation is lost.

\subsection{Real-Valued Matrix Indices}

We then turn to using real numbers to encode matrix dimensions. The implementation is symmetric to the one using integers (Listing~\ref{lst:function}):

\begin{lstlisting}[language=caml]
type arg =
	| Rows
	| Cols
	| Value of real * real

type 'a t = arg -> 'a
\end{lstlisting}

\noindent A problem arises in recursive functions where matrix dimensions are used as decrementors in stopping conditions, for instance in the \lstinline{fold_rec} function used in the implementation of the folding operation.
\begin{lstlisting}[language=caml]
let rec fold_rec f cols i j (m: 'a t) =
	let dec i j =
		if j <=. 0. then (i-.1.,cols) else (i,j-.1.)
	in
	if (i <=. 0. && j <=. 0.) || (i <. 0. || j <. 0.) then (
		m (Value (i,j))
	) else (
		let i',j' = dec i j in
		f (m (Value (i,j))) (fold_rec f cols i' j' m)
	)

let fold (f : 'a -> 'b -> 'b) (m: 'a t) : 'b =
	let rows = m Rows -. 1. in
	let cols = m Cols -. 1. in
	fold_rec f cols rows cols m
\end{lstlisting}

\noindent Imandra only accepts definitions of functions for which it can prove termination. The dimensions being real numbers prevents Imandra from being able to prove termination without providing a custom measure. In order to define this measure, we need to connect the continuous world of reals with the discrete world of integers (and ultimately ordinals) for which we have induction principles. We chose to develop a \lstinline{floor} function that allows Imandra to prove termination with reals.

To prove termination of our \lstinline{fold_rec} function recursing along reals, we define an \lstinline{int_of_real : real -> int} function in Imandra, using a subsidiary \lstinline{floor : real -> int -> int} which computes an integer floor of a real by ``counting up'' using its integer argument. In fact, as matrices have non-negative dimensions, it suffices to only consider this conversion for non-negative reals, and we formalise only this. We then have to prove some subsidiary lemmas about the arithmetic of real-to-integer conversion, such as:
\begin{lstlisting}[language=caml]
lemma floor_mono x y b =
  Real.(x <= y && x >= 0. && y >= 0.)
  ==> floor x b <= floor y b
\end{lstlisting}

\begin{lstlisting}[language=caml]
lemma inc_by_one_bigger_conv x =
  Real.(x >= 0. ==> int_of_real (x + 1.0) > int_of_real x)
\end{lstlisting}

\noindent Armed with these results, we can then prove termination of \lstinline{fold_rec} and admit it into Imandra's logic via the ordinal pair measure below:
\begin{lstlisting}[language=caml]
[@@measure Ordinal.pair
            (Ordinal.of_int (int_of_real i))
            (Ordinal.of_int (int_of_real j))]
\end{lstlisting}

Extending the functional matrix implementation to reals was not trivial, but it did have a real payoff.
Using this representation, we were able to verify real-valued versions of the pruned ACAS Xu networks!
In both cases of integer and real-valued matrices, we pruned the networks to 10\% of their original size.
So, we still do not scale to the full ACAS Xu challenge.  However, the positive news is that the real-valued version
of the proofs
uses the same waterfall proof tactic of Imandra, and requires no extra effort from the programmer to complete the proof.
This result is significant bearing in mind that many functional and higher-order theorem provers are known to have significant drawbacks when switching to real numbers.

From the functional programming point of view, one may claim that this approach is not elegant enough because
it does not provide true polymorphism as it encodes matrix dimensions as reals.
This motivates us to try the third alternative, using \emph{records} with \emph{maps} to achieve polymorphism.

\subsection{Records}

Standard OCaml records are available in Imandra, though they do not support functions as fields. This is because all records are data values which must support a computable equality relation, and in general one cannot compute equality on functions. Internally in the logic, records correspond to algebraic data types with a single constructor and the record fields to named constructor arguments. Like product types, records allow us to group together values of different types, but with convenient accessors and update syntax based on field names, rather than position. This offers the possibility of polymorphism for our matrix type. 

The approach here is similar to the one in Section~\ref{sec:function}: matrices are stored as mappings between indices and values, which allows for constant-time access to the elements. However, instead of having the mapping be implemented as a function, here we implement it as a \lstinline{Map}, i.e. an unordered collection of (key;value) pairs where each key is unique, so that this ``payload'' can be included as the field of a record. 

\begin{lstlisting}[language=caml]
type 'a t = {
	rows: int;
	cols: int;
	vals: ((int*int), 'a) Map.t;
}
\end{lstlisting}

We can then use a convenient syntax to create a record of this type. For instance, a weights matrix from one of the ACAS Xu networks can be implemented as:

\begin{lstlisting}[language=caml]
let layer6_map =
	Map.add (0,10) (0.05374) @@
	Map.add (0,20) (0.05675) @@
	...
	Map.const 0.

let layer6_matrix = {
	rows = 5;
	cols = 51;
	vals = layer6_map;
}
\end{lstlisting}

Note that the matrix dimensions (and the underlying map's keys) are indeed encoded as integers, whereas the weights' values are reals. 

Similarly to the previous implementations, we define a number of useful matrix operations which will be used to define general neural network layer functions. For instance, the \lstinline{map2} function is defined thus:

\begin{lstlisting}[language=caml, label={lst:map2_records}]
let rec map2_rec (m: 'a t) (m': 'b t) (f: 'a -> 'b -> 'c) (cols: int) (i: int) (j: int) (res: ((int*int), 'c) Map.t): ((int*int), 'c) Map.t =
		let dec i j = 
			if j <= 0 then (i-1, cols) else (i,j-1)     
		in
		if i <= 0 && j <= 0 then (
			res
		) else (
			let (i',j') = dec i j in
			let new_value = f (nth m (i',j')) (nth m' (i', j')) in
			let res' = Map.add' res (i',j') new_value in
			map2_rec m m' f cols i' j' res'
		)
[@@adm i,j]

let map2 (f: 'a -> 'b -> 'c) (m: 'a t) (m': 'b t) : 'c t = 
	let rows = max (m.rows) (m'.rows) in
	let cols = max (m.cols) (m'.cols) in
	let vals = map2_rec m m' f cols rows cols (Map.const 0.) in
	{
		rows = rows;
		cols = cols;
		vals = vals;
	}
\end{lstlisting} 

Compared to the list implementation, this implementation has the benefit of providing constant-time access to matrix elements. However, compared to the implementation of matrices as functions, it uses recursion to iterate over matrix values which results in a high number of case-splits. This in turn results in lower scalability.
Compared to the previous section's results, none of the verification tests on pruned ACAS Xu benchmarks that terminated with the functional matrix implementation terminated with the records implementation.

Moreover, we can see in the above function definition that we lose considerable conciseness and readability.

In the end, the main interest of this implementation is its offering polymorphism. In all other regards, the functional implementation seems preferable.

\section{Conclusions}
Functional programming languages that are tightly coupled with automated reasoning capabilities, like Imandra, offer us the possibility to verify and perform inference with neural networks, which the library CheckINN aims to do. To that aim, implementing matrices and matrix operations is important.
We have shown different implementations of matrices and how each implementation influences verification in Imandra.

This study has three positive conclusions:
\begin{itemize}
\item Imandra's language is sufficiently flexible to give rise to implementations of several choices of matrix in the CheckINN library. Its proof heuristics adapt smoothly to these different implementations, with very little hints needed to figure out the appropriate proof strategy (induction, waterfall or SAT/SMT proving).

\item this flexibility bears benefits when it comes to diversifying the range of NN properties we verify: thus, matrices as lists made possible proofs of higher-order properties by induction, whereas matrices as functions were more amenable to automated proofs in SAT/SMT solving style.

\item the transition from integer-valued to real-valued NNs is possible in Imandra. This transition itself opens several choices for matrix representations. However, if the matrix representation is optimal for the task at hand, Imandra takes care of completing the proofs with reals and adapts its inductive waterfall method to the new data type automatically.
  This is a positive lesson to learn, as this is not always given in functional theorem provers.
\end{itemize}

The main drawback is our failure to scale to the full ACAS Xu problem regardless of the matrix implementation choice in CheckINN~\cite{DPKD22}. However, it may not come as a great surprise, as general-purpose SMT solvers do not scale to the problem, either.  It took  domain-specific algorithms such as \emph{ReluPlex} and special-purpose solvers such as \emph{Marabou}  to overcome the scaling problem~\cite{KaBaDiJuKo17Reluplex,KatzHIJLLSTWZDK19}.
This suggests future solutions that are somewhat orthogonal to the choice of the matrix representation:
\begin{itemize}
  \item interface with Marabou or other specialised NN solvers in order to scale;
\item work on a set of Imandra's native proof heuristics and tactics, tailored specifically to Imandra's NN formalisations.
\end{itemize}

In addition, evaluating CheckINN against other benchmarks would allow to assess more accurately its scalability on different problems, e.g. robustness verification of image classification networks on the MNIST dataset \cite{noauthor_vnn_nodate} or range analysis of randomly generated networks \cite{dutta_output_2018}. We leave these as future work. 

These conclusions provide a strong foundation to further develop the CheckINN library, as its aim is to offer verification of a wide array of neural network properties and we have shown that the choice of matrix implementation eventually influences the range of verifiable properties. 

Finally, we believe that the methods we described could be useful in other theorem provers (both first- and higher-order) that combine functional programming and automated proof methods, such as ACL2, PVS, Isabelle/HOL and Coq. 
For example, in all these systems functions defining matrix operations (e.g., convolution) over lists  are often more complex compared to their counterparts over matrices represented as functions, which can benefit from non-recursive definitions.
Overall, as these various prominent theorem proving systems work ultimately with functional programs over algebraic datatypes like Imandra, our core observations carry over to them in a natural way.

 %
%
%
\bibliographystyle{splncs04}
\bibliography{MatrixFormalisation}
\end{document}